\documentstyle[12pt]{article}

\input epsf

\newcommand{\be}{\begin{equation}}
\newcommand{\ee}{\end{equation}}
\newcommand{\bea}{\begin{eqnarray}}
\newcommand{\eea}{\end{eqnarray}}
\newcommand{\ba}{\begin{array}}
\newcommand{\ea}{\end{array}}
\newcommand{\reseteq}{\setcounter{equation}{0}}

\newcommand{\hsp}[1]{\hspace*{#1pt}}
\newcommand{\sij}{\sigma_{i}\sigma_{j}}

\newcommand{\nti}{\lim_{n\rightarrow\infty}}
\newcommand{\Ln}{\mbox{${\cal L}_{n}^{\mbox{}}$}}
\newcommand{\Lns}{\mbox{${\cal L}_{n}^{\ast}$}}
\newcommand{\sn}[1]{\mbox{sn($#1$)}}
\newcommand{\cn}[1]{\mbox{cn($#1$)}}
\newcommand{\dn}[1]{\mbox{dn($#1$)}}
\newcommand{\lhat}{\mbox{$\!\widehat{\, L\, }\!$}}
\newcommand{\rud}{\rule[-12pt]{0pt}{30pt}}

\newcommand{\zr}[1]{\mbox{\hspace*{#1em}}}
\newcommand{\RR}{\mbox{\zr{0.1}\rule{0.04em}{1.6ex}\zr{-0.05}{\sf R}}}

\newcommand{\ZZ}{\mbox{\sf Z\zr{-0.45}Z}}

\newcommand{\CMP}[3]{Commun.\ Math.\ Phys.\ {\bf #1} (19#2) #3}
\newcommand{\EPL}[3]{Europhys.\ Lett.\ {\bf #1} (19#2) #3}

\newcommand{\IJMPB}[3]{Int.\ J.\ Mod.\ Phys.\ {\bf B#1} (19#2) #3}

\newcommand{\JPhysA}[3]{J.\ Phys.\ {\bf A#1} (19#2) #3}
\newcommand{\JPhysC}[3]{J.\ Phys.\ {\bf C#1} (19#2) #3}
\newcommand{\JPhysFrance}[3]{J.\ Phys.\ France {\bf #1} (19#2) #3}
\newcommand{\JSP}[3]{J.\ Stat.\ Phys.\ {\bf #1} (19#2) #3}

\newcommand{\NPB}[3]{Nucl.\ Phys.\ {\bf B#1} (19#2) #3}
\newcommand{\PLA}[3]{Phys.\ Lett.\ {\bf A#1} (19#2) #3}
\newcommand{\PLB}[3]{Phys.\ Lett.\ {\bf B#1} (19#2) #3}

\newcommand{\PhysRev}[3]{Phys.\ Rev.\ {\bf #1} (19#2) #3}
\newcommand{\PRL}[3]{Phys.\ Rev.\ Lett.\ {\bf #1} (19#2) #3}
\newcommand{\PRSL}[3]{Proc.\ R.\ Soc.\ Lond.\ {\bf #1} (19#2) #3}
\newcommand{\PTRSL}[3]{Phil.\ Trans.\ R.\ Soc.\ Lond.\
                       {\bf #1} (19#2) #3}

\newcommand{\ZPhys}[3]{Z.\ Phys.\ {\bf #1} (19#2) #3}

\begin{document}


\begin{center}
{\LARGE\bf A Critical Ising Model \\[3mm]
 on the Labyrinth} \\[8mm]
{\large\sc 
  M.\ Baake$^{1}$, \hsp{1} U.\ Grimm$^{2}$ 
 \hsp{1} and \hsp{1}  R.\ J.\ Baxter$^{3}$} \\[4mm]
{\footnotesize \mbox{}\footnotemark[1]
 Institut f\"{u}r Theoretische Physik, Universit\"{a}t T\"{u}bingen,\\
        Auf der Morgenstelle 14, 72076 T\"{u}bingen, Germany} \\[2mm]
{\footnotesize \mbox{}\footnotemark[2]
 Instituut voor Theoretische Fysica, Universiteit van Amsterdam, \\
   Valckenierstraat 65, 1018 XE Amsterdam, The Netherlands} \\[2mm]
{\footnotesize \mbox{}\footnotemark[3]
   Mathematics and Theoretical Physics, IAS, \\
   Australian National University, Canberra ACT 0200, Australia} \\[8mm]
\end{center}


\begin{quote}
{\small\sf 
 A zero-field Ising model with ferromagnetic coupling constants
 on the so-called Labyrinth tiling is investigated. 
 Alternatively, this can be regarded as an Ising model on a 
 square lattice with a quasi-periodic distribution of up to
 eight different coupling constants. The
 duality transformation on this tiling is considered and
 the self-dual couplings are determined.
 Furthermore, we analyze the subclass of exactly solvable
 models in detail parametrizing the coupling constants 
 in terms of four rapidity parameters. 
 For those, the self-dual couplings correspond to the
 critical points which, as expected, belong to the Onsager universality
 class. 
}

\end{quote}



\renewcommand{\theequation}{\arabic{section}.\arabic{equation}}

\section{Introduction}
\reseteq

The understanding of phase transitions in 2D systems has considerably
increased during the last decade. In particular, a lot is known 
about the conformal structure at the critical point 
of a whole hierarchy of models \cite{Cardy}. 
On the other hand, numerous series of solvable 
vertex models \cite{Baxter,VertexModels} and 
IRF (Interaction-Round-a-Face) models 
\cite{Baxter,IRFModels} have been constructed.
Furthermore, the so-called chiral Potts models \cite{Vladi,Baxter88}
attracted a lot of interest, mainly due to the existence of
superintegrable cases \cite{Vladi} (see also \cite{Kedem} and references
therein) and applications to 3D systems \cite{BB93}. 
 
While many of these models can be solved in the
presence of a temperature-like deviation from criticality,
solvability is in general restricted to the zero-field case
(apart from e.g.\ the hard hexagon model \cite{Baxter} which is a 
limiting case of the triangular Ising model in a symmetry-breaking field).
Recently, a series of models which can be solved in the presence of a
symmetry breaking ``field'' was found 
\cite{DiluteModels}, among them a model
which belongs to the universality class of the Ising model
in a magnetic field.
But even on regular lattices, notably the square lattice,
the word ``solvable'' does not mean that all physically 
interesting quantities have been calculated explicitly,
in particular the computation of two-point (or even higher) 
correlation functions still appears to be a hopeless task
in the general case.

Even less, however, is known about models on non-periodic graphs.
The main reason is that quite a number of new phenomena
show up: already the Lee-Yang zeros of the classical 1D Ising model
on substitution chains reveal gap structures that are unknown in
the periodic case \cite{BaaGriPis94}. Similarly,
the Ising quantum chain on such substitution structures 
\cite{Benza89,LinTao92,Luck93,GriBaa94}
and the 2D Ising model with aperiodicity in
one direction \cite{Tracy88,Igloi93}
resemble the periodic situation
at its critical point only under special conditions. 
These examples also show that results from conformal field theory cannot
generally be applied to systems on non-regular graphs
which is not too surprising as they rely on the
basic assumption of conformal invariance of the system
at the critical point.

The restricted knowledge about non-periodic systems is due to the
significantly increased difficulty to reach rigorous results, the study
of 8-vertex models on relatively general grids \cite{Baxter78,Korepin} 
or certain IRF models on (rhombic) duals of regular de Bruijn grids
\cite{Korepin,Choy} being untypical exceptions (see \cite{deBruijn}
for a definition of the grid formalism).
Other investigations make use of various approximative techniques
such as, for instance,
real space renormalization group \cite{GordLuckOrl86,AoyOda87}, 
mean field calculations \cite{DoroSoka89},
computer simulations 
\cite{BhatHoJohn87,AmaAnaAth88,OkaNii88a,OkaNii88b,WilVau88},
series expansions \cite{AbeDot89},
and finite-size scaling methods \cite{Sor91}, but almost
exclusively deal with Ising or Potts models on Penrose tilings.

In this article, we consider the zero-field Ising model defined on
a different quasiperiodic graph in 2D.
We have chosen the so-called Labyrinth tiling,
see \cite{SireMossSad89} for a detailed geometric description. It turns
out that this tiling is self-dual as a graph. This allows 
-- under certain assumptions such as uniqueness of the critical point -- 
a simple argument to obtain information about the location of 
critical points, though it turns out to be incomplete.
In the case of the frequently studied Penrose patterns,
duality and correlation inequalities are less conclusive,
but have been used to obtain bounds
for the critical temperature \cite{BhatHoJohn87,OkaNii88b}.

{}From the Peierls argument \cite{Peierls,Griff,Ellis}
it is clear that there must exist
at least one phase transition in our system. We do not expect that
the continuous transition in the Ising model on regular lattices
becomes a first-order transition in our case since introducing
``randomness'' generally  tends to smooth out transitions, 
though it is certainly not completely obvious (e.g.,
the four-spin interaction Ising model on the tetrahedra of the
3D fcc lattice in fact 
has a first-order transition at its self-dual point \cite{3DIsing}).

Unfortunately, we could not decide the question of 
uniqueness of the critical point in the ferromagnetic case
in general, though
one might expect this to be true if all 
coupling constants are strictly positive
and uniformly bounded from below. 
To further attack this problem, we have investigated
the solvable cases of the model by means of the rapidity line approach
\cite{Baxter86}. We find a 4D manifold in the 8D space of 
couplings where the Ising model on the Labyrinth is solvable.
Therein, one finds a 3D critical surface, hence
it contains one temperature-like parameter that describes the
deviation from criticality.
The result is given explicitly
by means of some geometric properties of the Labyrinth.
In all these cases, the critical point is unique and belongs to the Onsager
universality class as expected \cite{AntKor88,Doro89}.

\section{The Labyrinth tiling}
\reseteq

To investigate exactly some typical properties of generic 
non-periodic graphs in 2D,
we need a simple example which is more complicated than the Cartesian product
of a Fibonacci chain and a 1D periodic lattice 
(which is what underlies the well-studied
Ising quantum chains), but still considerably simpler than a Penrose tiling.
The natural choice is the so-called Labyrinth tiling \cite{SireMossSad89},
which got its name from the properties of certain colourings.
There exist several ways to construct it, 
see \cite{SireMossSad89,Sire89} for a detailed description. 
For our purpose, an approach via the Cartesian product of a
1D substitution sequence with itself is most appropriate which we will now
briefly outline.

\subsection{The silver mean substitution rule}

Let us start with the generation of a 1D chain, 
the so-called silver mean chain,
by repeated application of the two-letter substitution rule
\be \label{subst}
  \varrho \, : \,\, \ba{rcl} a & \rightarrow & b \\
                             b & \rightarrow & bab \ea
\ee
to the letter $a$. The corresponding substitution matrix $R$ is obtained
from the standard abelianization process, compare 
\cite{BaaGriJos93} and refs.\ therein, and reads
\be \label{submat}
  R = \left( \ba{cc} 0 & 1 \\ 1 & 2 \ea \right) \, .
\ee
The Perron-Frobenius eigenvalue is the silver mean, 
$\lambda = 1 + \sqrt{2}$,
the corresponding eigenvector reads $(1,\lambda)^t$, or
$((3-\lambda)/2,(\lambda-1)/2)^t$ in statistical normalization. 
Since $R$ is symmetric, left and
right eigenvectors are the transpose of each other. 
Thus, the entries have
two meanings: on the one hand, 
they tell us the relative frequencies of $a$'s
and $b$'s in the infinite chain 
-- the ratio of which is the irrational
number $\lambda$ wherefore the chain 
cannot be periodic. On the other hand,
they give us the natural choice for 
the lengths of two intervals with which
we can represent the chain {\em geometrically} 
such that the substitution
rule (\ref{subst}) gives rise 
to a inflation/deflation symmetry.
In this setup (which we shall use for graphical presentation), $a$ 
stands for a short interval, while that attached to $b$ is a factor
of $\lambda$ longer.

\subsection{Construction of the Labyrinth}

The next step is to generate a 2D tiling. First, 
we take an orthogonal Cartesian product of two identical 
silver mean chains in the proper
geometric representation. This way, we obtain a quadrant with an
orthogonal grid, but two different spacings in each
direction. Now, starting from the lower left corner, we mark
every second vertex point of this grid and connect them with
the nearest neighbours of the same kind. This would then give a
fourfold symmetric, non-periodic tiling of the plane 
(after filling the 3 free quadrants
with rotated copies of the grid constructed).

In what follows, we shall, however, mainly need so-called periodic
approximants. They can easily be obtained from the periodic approximants
of the silver mean chain. The \mbox{$n$-th} 
iteration of the substitution
rule (remember that we start with the letter $a$ for $n=0$) 
gives a finite chain with $g_{n}$ cells with the generalized
Fibonacci numbers $g_0=1$, $g_1=1$, and
\mbox{$g_{n+1}=2 g_{n} + g_{n-1}$}, compare
eqs.~(4.19) and (4.20) in Ref.~\cite{BaaGriJos93}. 
Now, due to the special
choice of the substitution (\ref{subst}), we can close the chain
periodically not only after the full number of cells, but also
after $g_{n}-1$ cells, which is always an even number.

If we do that for our finite grid (obtained from the Cartesian
product of two identical finite chains), we wrap it on the torus
and get the periodic approximants we need, compare Fig.~1.
They have the nice property that neither new vertex configurations
nor other mismatches occur. Such an approximant, obtained from
the \mbox{$n$-th} iteration step of 
the substitution rule, is called \Ln .
Note that the number $N_n$ of vertices in \Ln\ is given by
\be
N_{n} \;\; = \;\; \frac{1}{2} (g_{n}-1)^{2} \, .
\label{e:vertnumb}
\ee

The tiling was constructed from half the points of the orthogonal grid.
If we use the other half, we 
obtain another tiling of the same kind, and, even more, the {\em
dual graph} to the Labyrinth tiling. This also establishes the
checkerboard structure we will need later.
The duality of the two graphs remains true for the
periodic approximants, and Fig.~2 shows both \Ln\
and \Lns\ for $n=4$.
It is also obvious that the Labyrinth tiling is topologically a
square tiling -- but we have now performed a non-periodic deformation.
Furthermore, this tiling is quasiperiodic \cite{SireMossSad89,Sire89},
but it can be seen as a modulated structure and is therefore not
a true example of a quasicrystal \cite{Katz}. 
But this does not matter for our purposes 
as many arguments, in particular the rapidity line approach of
Sec.~4, apply also to discrete, locally finite structures obtained
from regular grids via dualization, compare 
\cite{Baxter78,Korepin,Choy}. Note, however, that 
these examples are usually not self-dual as a graph.

Before we proceed, let us describe some properties of the Labyrinth
tiling which are useful in our present context, for additional material
we refer to \cite{SireMossSad89,Sire89}. First of all, the tiling is
built from three tiles, a square (abbrev.\ by the letter $A$ from now on), 
a kite ($B$) and a trapezoid ($C$). They show
altogether three different edge lengths which we call $\ell$, $m$,
and $s$ for long, medium, and short, respectively.
We will now summarize their statistics.

\subsection{Tiles and vertices}

The substitution rule (\ref{subst}) of the underlying 1D chain induces a 
substitution rule for the 2D tiles which is local, compare
Fig.\ 3. The combinatorial net rule reads
\be \label{subst1}
 \ba{rcl}
    A   &  \rightarrow  &  A + 4 B + 4 C   \\
    B   &  \rightarrow  &  A + 3 B + 2 C   \\
    C   &  \rightarrow  &  A + 2 B + C \, .
  \ea    
\ee
This rule can be summarized by the tile substitution matrix $R_t$
\be \label{submat2}
   R_t = \left( \ba{ccc} 1 & 1 & 1 \\ 4 & 3 & 2 \\
                         4 & 2 & 1 \ea \right) \, ,
\ee
compare fig.~2(b) of \cite{SireMossSad89}. Since this matrix is no longer
symmetric, one has to distinguish geometric and statistical quantities.
We have chosen the convention where the statistical properties can be
read from the usual (right) eigenvectors.

The Perron-Frobenius eigenvalue is 
$\lambda^2 = 3 + 2 \sqrt{2}$ with eigenvector 
$(5 - 2 \lambda, 6 \lambda - 14, 10 - 4 \lambda)^t$
which shows the frequencies of the tiles in the thermodynamic limit.
Due to self-duality of the tiling as graph, this tells us also the
frequencies of the vertex configurations, see Fig.\ 4. Here, the four
orientations of the kite and the trapezoid (resp.\ that of the corresponding
vertex configurations) are summed over, respectively. Since the tiling has
fourfold symmetry \cite{SireMossSad89}, 
each orientation is equally frequent which we will use later.

\subsection{Statistics of bonds}

For the Ising model to be studied next, we need some information about
the {\em bonds} of the tiling. One important observation, compare
Figs.\ 1 and 2, is that transversally intersecting dual bonds of ${\cal L}$ and
${\cal L}^*$ are always of the same type (i.e., $\ell$ intersects with
$\ell^*$ etc.). Furthermore, this duality is still consistent with an
anisotropic generalization, where we distinguish altogether eight different
types of bonds. This holds true both for the
thermodynamic limit and for the finite approximants.

Also, we can calculate the frequency
of the bonds. The substitution rule for the tiles, 
summarized in (\ref{submat2}),
induces the following edge replacement rule,
\be \label{subst2}
  \ba{rcl}
    \ell         &  \rightarrow  &  4 \ell + 4 m + s   \\
   \ell + m      &  \rightarrow  &  6 \ell + 5 m + s   \\
 \ell + 2 m + s  &  \rightarrow  &  9 \ell + 6 m + s \, .
  \ea    
\ee
{}From this, one can read $m \rightarrow 2 \ell + m$ and
$s \rightarrow \ell$, wherefore the corresponding bond substitution
matrix $R_b$ reads
\be \label{submat3}
   R_b = \left( \ba{ccc} 4 & 2 & 1 \\ 4 & 1 & 0 \\
                         1 & 0 & 0 \ea \right) \, .
\ee
Here, the Perron-Frobenius eigenvalue is again $\lambda$, now with 
eigenvector $(\frac{1}{2}, \lambda -2, \frac{5}{2} - \lambda)^t$ in
statistical normalization, i.e., exactly 50\% of the bonds in the
thermodynamic limit are of type $\ell$ etc.
Again, different orientations of the same type of bond are summed over,
but are equally frequent due to the symmetry of the tiling.
We have now all geometric
data we need for the description of the Ising model.

\section{Ising model and duality}
\reseteq

We now consider a zero-field Ising model
with spins $\sigma\in\{1,-1\}$ located on the
vertices of the periodic approximant \Ln.
Let $J_{\ell}$, $J_{m}$, $J_{s}$ denote ferromagnetic
coupling constants for
neighbouring spins connected by a long,
medium, or short bond, respectively. In other
words, a bond of type $x$ between neighbouring
spins at locations $i$ and $j$ contributes 
$-J_{x}\sij$ to the total energy,
where $J_{x}\geq 0$ and $x\in\{\ell,m,s\}$. 

The canonical partition function $Z_{n}$ of the periodic approximant
\Ln\ is the sum over all configurations $\sigma$ on \Ln\
and has the form
\be
Z_{n} \;\; = \;\; \sum_{\sigma}
\left(
\prod_{<i,j>_{\ell}} \exp(K_{\ell}^{\mbox{}}\sij ) 
\prod_{<i,j>_{m}} \exp(K_{m}^{\mbox{}}\sij )
\prod_{<i,j>_{s}} \exp(K_{s}^{\mbox{}}\sij ) \right) \, .
\label{e:pf}
\ee
Here, $<\! i,j \! >_{x}$ denotes all nearest neighbour
pairs at positions $i$ and $j$
which are connected by a bond of type $x$ and
$K_{x}^{\mbox{}}=\beta J_{x}$, $x\in\{\ell,m,s\}$,
where $\beta=1/k_{B}T$ is the inverse temperature.  

In complete analogy to the treatment of the
square-lattice Ising model in \cite{Baxter} (we 
use the same notation throughout the argument)
we can write down both low- and high-temperature
expansions for the partition function (\ref{e:pf})
in terms of polygons on the lattice \Ln .
Let us commence with the high-temperature series.

%
%

\subsection{High-temperature expansion}

Consider an arbitrary, but fixed \Ln.
Since $\sij\in\{-1,1\}$, one has
\be
\exp(K\sij )\;\; = \;\;\cosh(K) 
\left( 1 + \tanh(K) \sij \right) \, .
\ee
Now, we introduce, for $x\in\{\ell,m,s\}$, 
\be
v_{x}^{\mbox{}} \;\; = \;\; \tanh (K_{x}^{\mbox{}})
\ee 
insert this and the previous expression into
(\ref{e:pf}), and expand the products.
If $m_x$ denotes the number of bonds of type $x$ in \Ln , 
we obtain $2^{m_{\ell}+m_m+m_s}$ terms of the form
\be \label{s-term}
   v_{\ell}^{r_{\ell}} v_{m}^{r_{m}} v_{s}^{r_{s}}
   \cdot \prod_{i=1}^{N_n} \sigma_i^{\nu_i} \; ,
\ee 
where $r_x$ is the total number of $x$-bonds in the polygonal
representation of the corresponding configuration, compare \cite{Baxter},
and $\nu_i$ is the number of lines with site $i$ as an endpoint.
Since $\sigma_i \in \{+1,-1\}$, summation of (\ref{s-term}) over the spin 
configurations will not contribute to $Z_n$ unless all $\nu_i$ are even,
when it sums up to
$v_{\ell}^{r_{\ell}} v_{m}^{r_{m}} v_{s}^{r_{s}} \cdot 2^{N_n}$.

This allows us to rewrite the partition function (\ref{e:pf})
as follows \cite{Baxter}
\be 
Z_{n} \;\; = \;\;
2^{N_{n}} {(\cosh K_{\ell}^{\mbox{}})}^{m_{\ell}}
{(\cosh K_{m}^{\mbox{}})}^{m_{m}}
{(\cosh K_{s}^{\mbox{}})}^{m_{s}}
\sum_{P\subset\Ln} 
v_{\ell}^{r_{\ell}} v_{m}^{r_{m}} v_{s}^{r_{s}}
\label{e:hte}
\ee
where the summation is now performed over all 
{\em polygon configurations} $P$ on \Ln . 
Note that (\ref{e:hte}) is the exact expression for any
(finite) periodic approximant because we do not truncate the series --
the name only stems from the importance ordering w.r.t.\ high temperature.
We will now turn to the appropriate counterpart for low temperature.

%
%

\subsection{Low-temperature expansion}

Let $m_x$ be as above and denote by $r_{x}$ the number of {\em unlike}
nearest-neighbour spins linked by a bond of type $x$. Then, for a given
configuration, there remain 
$m_{\ell}-r_{\ell}$ 
spin pairs of type $\ell$ 
etc., wherefore the corresponding term in $Z_{n}$ takes the value
\be
\exp(K_{\ell}^{\mbox{}} (m_{\ell} - 2 r_{\ell}) + 
K_m^{\mbox{}} (m_m - 2 r_m) +
K_s^{\mbox{}} (m_s - 2 r_s) ) \; .
\ee
Now, if any two adjacent spins in \Ln\ are different (and only then), 
we draw the {\em dual} bond in \Lns\ which is of the same type.
This gives us a set of precisely $r_{\ell},r_m,r_s$ lines (bonds)
of type $\ell,m,s$ in \Lns, respectively, which form closed polygons.

Since, in turn, each such polygon configuration represents exactly two
spin configurations, the partition sum (\ref{e:pf}) can be
rewritten as \cite{Baxter}
\be
Z_{n} \;\; = \;\;
2 \:\exp\left(
K_{\ell}^{\mbox{}} m_{\ell}+K_{m}^{\mbox{}} m_{m}+
K_{s}^{\mbox{}} m_{s}\right)
\sum_{P\subset\Lns}
\exp\left(
-2K_{\ell}^{\mbox{}} r_{\ell}-
2K_{m}^{\mbox{}} r_{m}-2K_{s}^{\mbox{}} r_{s}\right) \; .
\label{e:lte}
\ee
The sum is now over all polygon configurations
$P$ in the dual lattice \Lns ,
which, however, is isomorphic with the lattice \Ln\
itself wherefore the sum over \Lns\ can be replaced by one over \Ln.
Then, we can directly compare with Eq.~(\ref{e:hte}) above because
the meaning of $m_{\ell},r_{\ell}$ etc.\ is now the same.

Note that there is a slight
difference from the square-lattice model in \cite{Baxter}
where the horizontal and vertical lines interchange by
going to the dual lattice. In our case above, we 
only associated different coupling constants to bonds of
different length. We will discuss an anisotropic generalization
with eight different (ferromagnetic) coupling constants later.

%
%

\subsection{Free energy}

The average free energy per site $f$,
respectively the corresponding dimensionless
quantity $\psi$,
\be
\beta f\;\; = \;\; \psi \;\; = \;\;
- \nti \frac{1}{N_{n}}\ln Z_{n}
\ee
can now be expressed in two ways using the
above expansions (\ref{e:hte}) and (\ref{e:lte})
\bea
-\psi & = &
K_{\ell}^{\mbox{}} \mu_{\ell} \; +\; 
K_{m}^{\mbox{}} \mu_{m}\; +\; 
K_{s}^{\mbox{}} \mu_{s} \; +\; 
\Phi(e^{-2K_{\ell}^{\mbox{}}},e^{-2K_{m}^{\mbox{}}},e^{-2K_{s}^{\mbox{}}}) 
\label{free1} \\
& = &
\ln\left[ 2\, {(\cosh K_{\ell}^{\mbox{}})}^{\mu_{\ell}}\,
{(\cosh K_{m}^{\mbox{}})}^{\mu_{m}}\,
{(\cosh K_{s}^{\mbox{}})}^{\mu_{s}}\right] 
\; +\; \Phi(v_{\ell}^{\mbox{}},v_{m}^{\mbox{}},v_{s}^{\mbox{}}) \; .
\label{free2}
\eea
Here, we introduced the notation
\be
\mu_{x} \;\; = \;\; \nti \frac{m_{x}(n)}{N_{n}}
\hspace*{10mm} (x\in\{\ell,m,s\})
\label{e:mud}
\ee
and
\be
\Phi(v_{\ell}^{\mbox{}},v_{m}^{\mbox{}},v_{s}^{\mbox{}}) \;\; =\;\;
\nti \frac{1}{N_{n}} \ln\left(\sum_{P\subset\Ln}
v_{\ell}^{r_{\ell}} v_{m}^{r_{m}} v_{s}^{r_{s}}\right)
\ee
Thus $\mu_{x}$ is the frequency of the bonds
of type $x$ relative to the number of vertices, 
which means that the $\mu_{x}$ are normalized according to
\be
\mu_{\ell} + \mu_{m} + \mu_{s} \;\; = \;\; 2 \; .
\label{e:mus}
\ee
This follows from the co-ordination number of any vertex being $n_{c}=4$ and each bond
belonging to precisely two vertices. 
The actual values follow from the remarks after Eq.~(\ref{submat3}) and 
are given by
\be
\mu_{\ell} \; =\; 1 \; , \hspace*{8mm}
\mu_{m} \; =\; 2\lambda -4 \; , \hspace*{8mm}
\mu_{s} \; =\; 5-2\lambda 
\label{e:muv}
\ee
with $\lambda=1+\sqrt{2}$.

%
%

\subsection{Duality and critical point}

We introduce dual couplings $K_{x}^{\ast}$ by
\be
\exp (-2K_{x}^{\mbox{}}) \;\; = \;\; \tanh(K_{x}^{\ast})
\;\; = \;\; v_{x}^{\ast}
\label{dualcoup}
\ee
(or, equivalently, by
\mbox{$\exp(-2K_{x}^{\ast})=\tanh(K_{x}^{\mbox{}})$})
with $x\in\{\ell,m,s\}$. They satisfy
\be
\sinh (2K_{x}^{\mbox{}}) \: \sinh (2K_{x}^{\ast})
\;\;= \;\; 1
\label{dualcoup1}
\ee
which is a more symmetric way to present the
relationship between the two sets of couplings.
Also, Eq.~(\ref{dualcoup1}) shows directly that the
duality transformation 
\be \label{dualtra1}
     * \; : \;\;\; K_{x}^{ } \; \longmapsto \; K_{x}^{\ast}
\ee
is an involution on $\RR_+^3$, where it is also analytic. 
Using Eq.~(\ref{e:mus}),
one obtains from Eqs.\ (\ref{free1}) and (\ref{free2})
the following transformation of $\psi$ under $*$,
\be
\psi (K_{\ell}^{\ast},K_{m}^{\ast},K_{s}^{\ast})
\;\; = \;\; \psi(K_{\ell}^{\mbox{}},K_{m}^{\mbox{}},K_{s}^{\mbox{}}) \; +\;
\frac{1}{2} \ln\left[
{(\sinh 2K_{\ell}^{\mbox{}})}^{\mu_{\ell}}
{(\sinh 2K_{m}^{\mbox{}})}^{\mu_{m}}
{(\sinh 2K_{s}^{\mbox{}})}^{\mu_{s}}
\right] \; .
\label{e:dual}
\ee

%
%

Therefore, the duality transformation relates the free energies
at different couplings which implicitly depend on temperature.
In general, they do not belong to the same Ising system, wherefore
we cannot say much about phase transition points
($=$ non-analyticity points of $\psi$). However, if we
define
\be \label{invset}
    {\cal M} \; := \;
    \{ (K_{\ell},K_m,K_s) \in \RR_+^3 \; | \; 
       \psi(K_{\ell},K_m,K_s) \mbox{ not analytic } \}
\ee
we know that ${\cal M}$ is invariant under duality transformation $*$.
It would now be interesting to know the hierarchy of $*$-invariant
subsets of ${\cal M}$ because this would tell us the critical structure
of the model, but, unfortunately, we do not have enough information
about ${\cal M}$ to do so. Let us therefore at least look for
invariant points.
The only fixed points of Eq.~(\ref{dualtra1}) are obtained from
\be
 \sinh(2K_{\ell}^{\mbox{}}) \; = \;
 \sinh(2K_{m}^{\mbox{}}) \; = \;
 \sinh(2K_{s}^{\mbox{}}) \; = \; 1 \; ,
 \label{fixpo}
\ee
i.e., 
\mbox{$K_{\ell}^{\mbox{}}=K_{m}^{\mbox{}}=K_{s}^{\mbox{}}=
\ln(1\! +\!\sqrt{2})/2=0.44068679\ldots$}.
This, of course, is just the phase transition point of the periodic
square lattice Ising model \cite{KraWa,Baxter}. Unfortunately,
this does not tell us anything about the critical behaviour
in the non-periodic case.

Before we continue, let us remark that 
the additional term on the right-hand side
of Eq.~(\ref{e:dual}), which is anti-symmetric under duality
transformation, vanishes at the self-dual point. 
Therefore, the latter lies on the surface defined by
\be
{(\sinh 2K_{\ell}^{\mbox{}})}^{\mu_{\ell}}
{(\sinh 2K_{m}^{\mbox{}})}^{\mu_{m}}
{(\sinh 2K_{s}^{\mbox{}})}^{\mu_{s}}
\;\; = \;\; 1
\label{surface}
\ee
with $\mu_{\ell}$, $\mu_{m}$, and
$\mu_{s}$ as listed in Eq.~(\ref{e:muv}).
This surface consists of all points in coupling space 
for which the free energy is invariant under duality transformation
(though the couplings may change). 
It would be interesting to see how 
phase transition points for non-periodic models
(which certainly exist for non-vanishing couplings
 due to the Peierls argument, see \cite{Peierls}, 
 ch.~V of \cite{Griff} and, for the proper generalization needed here,
 ch.~V.5 of \cite{Ellis})
are located relative to this surface, i.e., how ${\cal M}$
is related to it.

As mentioned before, we do not know the full solution for the
non-periodic model. In the ferromagnetic regime, one might
expect a single critical point of the Onsager universality class, 
but even this appears to be quite complicated to decide.

If one also allows anti-ferromagnetic interactions,
the general picture will, of course, be different due to
the presence of frustration, compare~\cite{DuneDunlOgu93}
(see also \cite{Luck87,Sire93} for the effects of
frustration even in the 1D case).  
Things get also more complicated then by local energetic
degeneracies, a whole hierarchy of which can be seen as a
consequence of inflation/deflation symmetries 
\cite{OitAydJohn90,DuneDunlOgu93}.
However, if {\em all} interactions are anti-ferromagnetic,
the behaviour of the system will be the same as in the
purely ferromagnetic case since the Labyrinth is a
bipartite graph and an overall sign
can be absorbed reversing the spins on one sublattice. 

\subsection{Anisotropic generalization}

So far, we have dealt with three types of bonds only. Like in the 
square lattice case, compare \cite{Baxter}, we can however distinguish 
``raising'' bonds from ``lowering'' bonds. If we introduce the label
$xy$ for the box with abscissa $x$ and ordinate $y$ and 
\mbox{$K_{xy}^{\mbox{}}=\beta J_{xy}^{+}$}
\mbox{($L_{xy}^{\mbox{}}=\beta J_{xy}^{-}$)} 
for the couplings of a raising (lowering) bond in it
(see Fig.\ 5),
it is easy to derive from Fig.\ 2 that the proper duality 
transformation\footnote{We will use the same symbol since
misunderstandings are unlikely.} is now given by
\be \label{gendual}
\ast\; : \hspace*{10mm}
\left( \ba{c} K_{xy}^{\mbox{}} \\ L_{xy}^{\mbox{}} \ea \right)
 \;\; \longmapsto\;\;
\left( \ba{c} K^{\ast}_{xy} \\ L^{\ast}_{xy} \ea \right)
\ee
with $K^{\ast}_{xy}$ and $L^{\ast}_{xy}$ defined by
\be \label{gendual1}
\ba{rcl}
\sinh(2K^{\ast}_{xy})\, \sinh(2L_{xy}^{\mbox{}}) & = & 1 \\
\sinh(2L^{\ast}_{xy})\, \sinh(2K_{xy}^{\mbox{}}) & = & 1 \; .
\ea \ee
Here, $xy$ is any element of the set of labels $\{aa,ab,ba,bb\}$
where $a$ and $b$ refer to (\ref{subst}) and hence to a short and
a long interval, respectively, of the original 1D chain, see Fig.\ 1.
Again, $*$ is an analytic involution, this time on $\RR_+^8$.

Also, one can use the definition (\ref{invset}) w.r.t.\ this
8D space. From the Peierls argument, one would expect the set
${\cal M}$ to be some sort of codimension one manifold, possibly
with a rather complicated structure, where again the knowledge of
$*$-invariant subsets is desirable. Let us look at the submanifold 
${\cal S}$ of self-dual points, which is easily calculable. With
the abbreviation
\be \label{sinhyp}
  S_{xy} := \sinh(2 K_{xy}^{\mbox{}})\, \sinh(2 L_{xy}^{\mbox{}})
\ee
one obtains the fixed points of Eq.~(\ref{gendual}) as the
solutions of
\be \label{critsurf}
S_{aa}\; =\; S_{ab}\; =\; S_{ba}\; =\; S_{bb}\; =\; 1 \; . 
\ee
A little later, we will relate this self-duality surface ${\cal S}$ to
critical points obtained from exactly solvable cases.

In order to identify the 4D manifold ${\cal S}$
defined by Eq.~(\ref{critsurf}) with a surface of critical points,
one needs the following arguments and assumptions.
For every point $\xi\in{\cal S}$ one can easily find a curve 
$K_{xy}(\tau)$, $L_{xy}(\tau)$ parametrized by $\tau\in\RR_+$ 
with the following properties: 
(i)   $K_{xy}(\tau)$ and $L_{xy}(\tau)$ are strictly 
      increasing functions of $\tau$, 
(ii)  the curve is mapped onto itself under duality, 
(iii) for some (unique) $\tau_0$, the curve passes through $\xi$,
(iv)  $K_{xy}(0)=L_{xy}(0)=0$ and $K_{xy}(\infty)=L_{xy}(\infty)=\infty$.
Obviously, such a curve need not belong to a single Ising 
model\footnote{Remember that this is defined through a 
{\em fixed} set of coupling constants $J_{xy}^{\pm}$.}. Nevertheless,
$\tau$ plays the role of an inverse temperature wherefore the usual
arguments apply: 
(a) there is at least one phase transition along the curve, 
    compare \cite{Peierls}, and
(b) assuming uniqueness, the phase transition point coincides with the
    self-dual point $\xi$.
We shall say more about criticality in Sec.~4.

Of course, one can also work, step by step, through the generalizations of the
equations of the previous sections to find the transformation of the
free energy under $*$.
The manifold of couplings where the free energy is
invariant under duality is now given by
\be \label{critsurf2}
  (S_{bb})^{\mu_{\ell}/2}\, (S_{ba})^{\mu_m/4}\,
  (S_{ab})^{\mu_m/4}\, (S_{aa})^{\mu_s/2} \; = \; 1 
\ee
which contains ${\cal S}$ as a submanifold.
In Eq.~(\ref{critsurf2}), 
we have used the frequencies of the bonds and their isotropic
distribution over the different orientations. 

Again, it would be
instructive to see how the various surfaces are interrelated,
which we will now partially explore for a subspace of
coupling space where the model is exactly solvable. Here, the
set of critical points in fact coincides with the corresponding
submanifold of the self-duality surface.

\section{Rapidity lines and solvability}
\reseteq

The checkerboard structure of the original graph gives us the opportunity
to introduce so-called rapidity lines for the parametrization of the
couplings, compare \cite{Baxter86} for details.
Whereas Baxter considered the most general case of a ``$Z$-invariant''
Ising model \cite{Baxter86} (which is a special case of the 
exactly solvable ``$Z$-invariant'' zero-field eight-vertex 
model \cite{Baxter78}), 
we are only interested in models which have
at most eight different coupling constants associated with the
eight different bonds in the Labyrinth tiling.
This leads to exactly solvable models with commuting transfer matrices
and couplings $K_{xy}$ and $L_{xy}$ parametrized by 
differences of four rapidity parameters 
$u_{x}^{\mbox{}}$ and  $u^{\prime}_{y}$
(where both $x$ and $y$ can be either $a$ or $b$)
for the horizontal and vertical rapidity lines, respectively,
and one temperature-like variable $\Omega$ which describes 
the deviation from criticality.
Altogether, this means that we have a 4D solvable
subspace of the 8D space of couplings
$K_{xy}$ and $L_{xy}$ (which, of course, implicitly contain the 
temperature).

Fig.\ 6 shows a small section of the Labyrinth tiling with the
underlying grid (light) and the rapidity lines (dashed) with the
associated parameters. Here, the rapidity lines are constructed
parallel to the underlying grid in such way that they divide
the edges of the underlying grid into two equal halves.
Note that this gives a one-to-one 
correspondence between the intersections of two rapidity
lines and the bonds of the tiling on which the intersection
occurs. The checkerboard structure is apparent as the vertices
of the Labyrinth lie in rectangles which constitute one of the two
sublattices of the rectangular grid formed by the rapidity lines.

It is not possible to give a full account of the actual solution in what
follows. Instead, with special focus on the Labyrinth,
we summarize the key points of the approach of Ref.\ 
\cite{Baxter86} which the interested reader should have at
hand for cross-reference. Though we can simplify
several of the conditions (and actually give the results without derivation)
we still depend on the explicit parametrization by elliptic functions which
we will now outline. 

\subsection{Elliptic parametrization and duality}

One can explicitly obtain parametrizations of the 
couplings in the integrable case 
for three different regimes \cite{Baxter86}, namely
\be
\sinh (2K_{xy}^{\mbox{}}) \; = \; F(\eta-u^{\prime}_{x}+u_{y}^{\mbox{}}) \; ,
\hspace{10mm} 
\sinh (2L_{xy}^{\mbox{}}) \; = \; F(u^{\prime}_{x}-u_{y}^{\mbox{}}) 
\label{param}
\ee
where $F(u)$ is a quotient of Jacobi's elliptic functions 
$\sn{u}$, $\cn{u}$, and $\dn{u}$ and where
$\eta$ is determined by their half-period magnitudes $K$ and
$K^{\prime}$, see Table~1. The rapidities $u_{x}^{\mbox{}}$ 
and $u^{\prime}_{x}$ ($x\in\{a,b\}$) are a priori arbitrary complex numbers,
but in order to obtain real (positive) coupling constants
they have to be chosen appropriately, see Ref.~\cite{Baxter86}
for details. 
In Table~1, we also define several quantities
which will be used in what follows. Here, $k$ and $k^{\prime}$
denote the elliptic modulus and conjugate modulus,
respectively, and $\Theta(u)$ and $\Theta_{1}(u)$ are the usual
Jacobian theta functions \cite{GradRyz65}.

\vspace{10mm}
\begin{footnotesize}
\begin{center}
{\small{\bf Table 1:}\hspace*{2mm} 
 Parametrization in terms of elliptic functions}\\*[4mm]
\begin{tabular}{|c|c|c|c|c|}
\hline
 & regime I \rud & criticality  & 
 regime II    & regime III \\
\hline
 $\Omega$ \rud & {\large $\frac{1}{k^{\prime}}$}  & $1$    & 
 $k^{\prime}$  & {\large $\frac{ik^{\prime}}{k}$} \\
$\eta$ \rud    & $K$             & {\large $\frac{\pi}{2}$}    & 
 $K$           & $K-iK^{\prime}$ \\ 
$F(u)$ \rud    & $\frac{\sn{u}}{\cn{u}}$   & $\tan(u)$   &
 $k^{\prime}\frac{\sn{u}}{\cn{u}}$  & $ik^{\prime}\frac{\sn{u}}{\cn{u}}$ \\
$r(u)$\rud    & $\frac{\dn{u}}{\sn{u}\cn{u}}$ & 
 {\large $\frac{2}{\sin(2u)}$} &
 $\frac{\dn{u}}{\sn{u}\cn{u}}$ & $\frac{\cn{u}}{\sn{u}\dn{u}}$ \\
$q(u)$ \rud   & {\large
 $\frac{K^{\prime}}{\pi}\frac{\Theta_{1}^{\prime}(u)}{\Theta_{1}(u)}$} & 
 $0$ & {\large
 $\frac{K^{\prime}}{\pi}\frac{\Theta^{\prime}(u)}{\Theta(u)}$} & {\large
 $\frac{K^{\prime}}{\pi}\frac{\Theta^{\prime}(u)}{\Theta(u)}$} \\
$c$   \rud    & {\large $\frac{k^2 K^{\prime}}{\pi}$ } & $0$ &
 {\large $-\frac{k^2 K^{\prime}}{\pi}$} & 
 {\large $-\frac{K^{\prime}}{\pi}$} \\
\hline
\end{tabular}
\end{center}
\end{footnotesize}
\vspace{10mm}

In all regimes, we require the elliptic moduli $k$, $k^{\prime}$ 
(\mbox{$k^{\prime}=\sqrt{1-k^2}$})
to satisfy the inequality \mbox{$0\leq k,k^{\prime}\leq 1$}.
Hence, the regimes are characterized by the values of $\Omega$ 
(which plays the role of a ``temperature-like'' variable) 
given in Table~1 as follows: on has
\mbox{$\Omega^{2}>1$}, \mbox{$0<\Omega^{2}<1$} and
\mbox{$\Omega^{2}<0$} for regimes I, II and III, respectively.
The critical surface (\mbox{$\Omega^2=1$}) separates  the
``low-temperature'' regime I from the disordered
``high-temperature'' regimes II and III.
At all other values of $\Omega$, the partition function turns out
to be analytic. We have listed regime III for the sake of completeness
because it might become interesting in the context of Lee-Yang zeros
in the complex temperature plane.

In the present context, we are mainly interested in regimes I and II
as these lead to positive real couplings for the
Ising model. In particular, this is the case if one chooses
real rapidity differences \mbox{$u^{\prime}_{x}-u_{y}^{\mbox{}}$}
within the range \mbox{$0<u^{\prime}_{x}-u_{y}^{\mbox{}}<\eta$}.
Obviously, these two regimes are also related by 
duality (which inverts $\Omega^2$) since the dual
couplings (defined by Eqs.~(\ref{gendual}) and (\ref{gendual1}))
are obtained from
\be
\sinh (2K_{xy}^{\ast})\; = \;\Omega^{-1} \, 
F(\eta-u^{\prime}_{x}+u_{y}^{\mbox{}}) \; ,
\hspace{10mm}
\sinh (2L_{xy}^{\ast})\; = \;\Omega^{-1} \, 
F(u^{\prime}_{x}-u_{y}^{\mbox{}}) \; .
\label{dualparam}
\ee
Note that in the parametrization given in Table~1
the critical case corresponds to the limits 
\mbox{$k\rightarrow 0$},
\mbox{$k^{\prime}\rightarrow 1$},
\mbox{$K\rightarrow \pi/2$},
\mbox{$K^{\prime}\rightarrow\infty$},
and \mbox{$q\rightarrow 0$} where 
\mbox{$q=\exp(-\pi K^{\prime}/K)$} is the
elliptic nome.

The above parametrization defines a subspace of 
couplings for which the Ising model is ``$Z$-invariant''.
The main ingredient in the derivation is, of
course, the star-triangle relation \cite{Baxter,Baxter86}.
Although we do not want to go into details of the calculation,
we still feel that a few words should be spent on the 
actual conditions on the couplings which are solved
by the parametrization (\ref{param}).

The restrictions imposed by the condition of
integrability can be conveniently summarized as
follows. To this end, it is convenient to introduce
another involution $\;\widehat{.}\;$,
\be\label{hut}
\widehat{.} \; : \hspace*{10mm} 
\left(\ba{c} K_{xy} \\ L_{xy}^{} \ea\right) 
\longmapsto 
\left(\ba{c} \widehat{K}_{xy} \\ \lhat_{xy} \ea\right)
\ee
where $\widehat{K}_{xy}$ and $\lhat_{xy}$ are implicitly given by
\be\label{hut1}
\sinh(2\widehat{K}_{xy})\, \sinh(2K_{xy}) \; =\; 1 \; ,\hspace*{10mm}
\sinh(2\lhat_{xy})\, \sinh(2L_{xy}) \; =\; 1 \; .
\ee
This involution, together with the duality transformation $\;\ast\;$,
generates the Abelian group \mbox{$\ZZ_2 \! \oplus \! \ZZ_2$}
(Klein's 4-group) which acts on the 8D coupling space $\RR_+^8$.
(Note that $\;\widehat{.}\;$ and $\;\ast\;$ 
coincided in Sec.~3.4 which was the reason for 
the ``poor'' self-duality condition). In the present context, 
integrability means that the equation
\be
\Omega^{2} \;\; = \;\;
\prod_{j=1}^{4} 
\frac{\sinh(2M_{j})\cosh(\widehat{M}_{1}+\widehat{M}_{2}+
\widehat{M}_{3}+\widehat{M}_{4}-2\widehat{M}_{j})}
{\sinh(2\widehat{M}_{j}) \cosh(M_{1}+M_{2}+M_{3}+M_{4}-2M_{j})}
\label{intcond}
\ee
has to be satisfied for all vertices, where $M_{j}$ 
($j=1,\ldots,4$) denote the couplings on the
four surrounding edges of the vertex.
In the Labyrinth, we have nine different vertices 
(counting orientations),
cf.\ Fig.~4, which yield five different equations (due to 
symmetries).
Note that the value of $\Omega$ is the same for all vertices,
hence one equation can be regarded as the definition of
$\Omega$ and the remaining four result in relations between
the couplings.
Using the notation of Eq.~(\ref{sinhyp}), four of the
five conditions can actually be simplified to
\be
S_{aa} \; = \; S_{ab} \; =\; S_{ba} \; = \; S_{bb} \; = \; \Omega \\[2mm]
\label{intcoup}
\ee
which is also apparent in the parametrization given in Table~1
as the function $F$ satisfies
\be
F(u)\:F(\eta-u)\;\; =\;\;\Omega 
\ee
in all three regimes. 

The one additional equation stems from the vertex which involves
one long, one short and two medium length bonds, see Fig.~4. 
The equivalence of the equations obtained from the four 
different orientations follows from Eq.~(\ref{intcoup}).

Note also that Eq.~(\ref{intcond}) is automatically satisfied
for vertex configurations obtained by including words
with sequences \mbox{$\ldots aa\ldots$} which do not occur in
the silver mean chain (\ref{subst}). This means that
the same conditions arise if one considers Ising models
on tilings which are defined by different
two-letter substitution rules in an analogous way.
If one wants to consider models which involve only the four conditions
of Eq.~(\ref{intcoup}), one is forced to use periodic sequences of
just one letter in one direction as this is the only way to avoid
the vertex which yields the additional equation. 
 
The conditions (\ref{intcond}) drastically restrict the
possible couplings for the integrable cases.  
In the ``isotropic'' case \mbox{$K_{aa}=L_{aa}=K_{s}$},
\mbox{$K_{ab}=L_{ab}=K_{ba}=L_{ba}=K_{m}$},
\mbox{$K_{bb}=L_{bb}=K_{\ell}$} considered in Sec.~3,
the solvable subspace reduces to the isotropic Ising model 
on the square lattice, since from Eq.~(\ref{intcoup}) one obtains
\mbox{$\sinh(2K_{s})=\sinh(2K_{m})=\sinh(2K_{\ell})=\Omega^{1/2}$}.

Of course, Eqs.~(\ref{intcond}) and (\ref{intcoup}) are entirely
consistent with the results of Sec.\ 3
as the intersection of the solvable subspace
with the self-dual surface  (\ref{critsurf2}) 
is just the {\em critical} solvable case with $\Omega=1$.
One should also realize that it is the additional equation
with respect to Eq.~(\ref{intcoup}) that actually restricts the
solvable critical manifold to a subspace of the self-dual
manifold in our case. If one considers models where this
equation is absent (which means that one uses a periodic sequence
$aaa...$ or $bbb...$ in one direction of the grid on which the tiling
is built), the two spaces coincide. 
It would be interesting to see explicitly how the extra
condition in the generic case affects the critical behaviour of the system. 

\subsection{Partition function and magnetization}

Provided that $\Omega$ and all couplings are real and positive,
we can follow the techniques outlined in 
\cite{Baxter78,Baxter86,Korepin,Choy}
to calculate the partition function in the thermodynamic limit. Here,
the result (normalized per site) reads 
\be
Z \;\; := \;\; \lim_{n\rightarrow\infty} Z_{n}^{1/N_{n}} 
\;\; = \;\; 
{(4\Omega)}^{1/4}\:\exp \left(
\frac{1}{2}\mu_{s}\xi_{aa} +
\frac{1}{4}\mu_{m}(\xi_{ab}+\xi_{ba}) +
\frac{1}{2}\mu_{\ell}\xi_{bb}\right)
\label{partfun}
\ee
where 
\be
\xi_{xy} \;\; = \;\;
\phi(u^{\prime}_{x}-u_{y}^{\mbox{}}) \; +\;
\phi(\eta-u^{\prime}_{x}+u_{y}^{\mbox{}})
\ee
and (cf.\ Table~1)
\be
\phi(u) \;\; = \;\;
\frac{1}{2}\, c\, u \; +\; 
\int_{0}^{u} \,\left(\frac{x}{2K} + q(x) \right)\, r(x)\, dx
\;\; .
\ee

The local spontaneous magnetization $\langle\sigma\rangle$
is given by \cite{Baxter86}
\be
\langle\sigma\rangle \;\; = \;\; 
\left\{ \ba{l@{\hspace*{8mm}}l}
{(1-\Omega^{-2})}^{1/8} & \mbox{if $\Omega^2>1$} \\
 0 & \mbox{if $\Omega^2\leq 1$} \ea \right.
\label{mag}
\ee
for {\em any} single Ising spin of the lattice --
wherefore we suppressed the site index in (\ref{mag}).
One can also easily find expressions for the 
correlations of two Ising spins that belong to the
same plaquette of the Labyrinth tiling \cite{Baxter86}.
These, of course, depend on the particular couplings along
the edges of that plaquette.

The above feature is certainly somewhat surprising as
one would expect the one-point function to depend on the
local neighbourhood. That this is not the case for the 
magnetization is clearly due to the conditions (\ref{intcoup})
on the couplings. It is to be expected that 
one would, for instance, observe a smaller magnetization
at the vertices which are surrounded by long bonds only 
if one chooses $K_{bb}$ and $L_{bb}$ to be small simultaneously
(compared to the other couplings) which
necessarily violates Eq.~(\ref{intcoup}).
In this sense, the integrable subspace is possibly not
representative and 
we presently cannot exclude the possibility of
more than one phase transition 
in the general case where the corresponding 
order parameters would be the magnetization of
certain subsets of the Labyrinth vertices.

%
%
%
%

\section{Concluding remarks}
\reseteq

A zero-field Ising model on the Labyrinth tiling has been investigated.
We used duality arguments to obtain information about the
location of critical points and determined the self-dual
couplings. Also, the subclass of exactly solvable models
was considered in detail, following closely the 
discussion of the checkerboard Ising model in Ref.~\cite{Baxter86}.
In this case, the self-dual couplings also define 
the critical surface, the transition being of the Onsager type
as in the periodic model. 

The special model investigated here is, of course,
just an example. It can be generalized to 
tilings obtained by an analogous procedure from
arbitrary two-letter substitution rules, provided one uses
the correct periodic approximants. In this way, one can
study Ising systems where the fluctuations of
the underlying tiling behave in different ways.
This is of interest because it has recently been shown in several 
closely related examples that these fluctuations actually determine the 
critical behaviour of the Ising model, see e.g.\
\cite{Tracy88,Luck93,GriBaa94,Igloi93}. 

Of course, the investigation of the solvable cases
can only be the first step in the investigation of the
critical properties of Ising models on aperiodic tilings,
and we hope to report on further results soon. 
The exactly solvable $Z$-invariant case is arguably too
restrictive to be assumed representative as it, for instance,
forces the magnetization to be independent of the location,
which will certainly not be the case for some arbitrarily
chosen couplings. 

Finally, one can extend the analysis to other systems like the more general
eight-vertex model in the spirit of \cite{Baxter78,Korepin,Choy}. 
Here, it is now relatively clear what happens in the solvable cases which are,
by no means, restricted to de Bruijn grids, cf.\ \cite{deBruijn}. 
However, any progress beyond {\em local} solvability 
would help understanding the hierarchy of critical phenomena
between periodic and random.

%
%
%
%

\section*{Acknowledgements}

It is a pleasure to thank B.\ Nienhuis and P.\ A.\ Pearce
for valuable discussions. 
M.\ B.\ would like to thank P.\ A.\ Pearce 
and the Department of Mathematics,
University of Melbourne, for hospitality, 
where part of this work was done.
This work was supported by Deutsche Forschungsgemeinschaft (DFG)
and Stichting voor Fundamenteel Onderzoek der Materie (FOM).



\clearpage

\begin{figure}
\centerline{\epsfxsize=\textwidth\epsfbox[72 160 540 630]{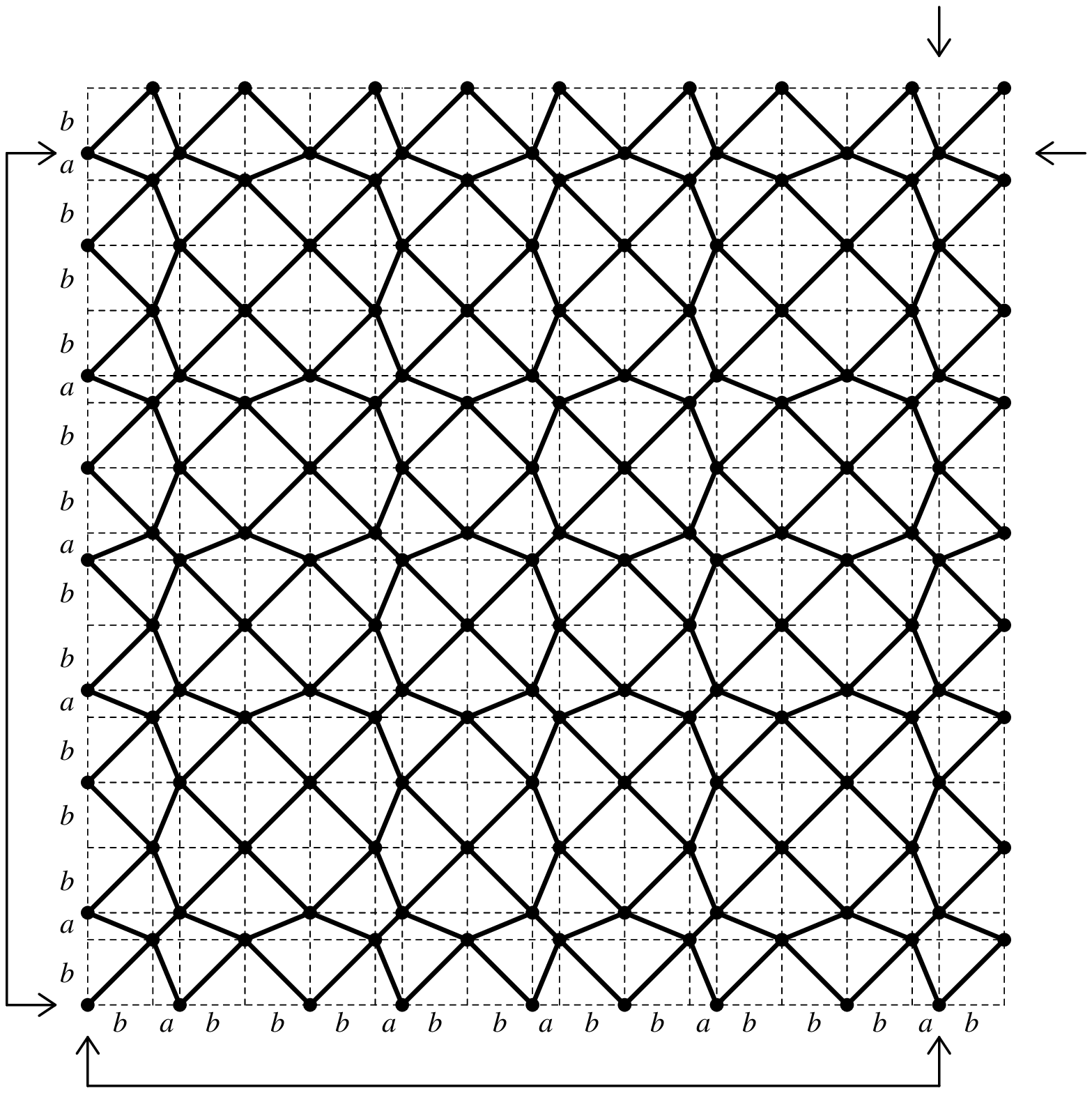}}
\caption{The Labyrinth with underlying grid and boundary conditions}
\end{figure}\clearpage

\begin{figure}
\centerline{\epsfxsize=\textwidth\epsfbox[72 160 540 630]{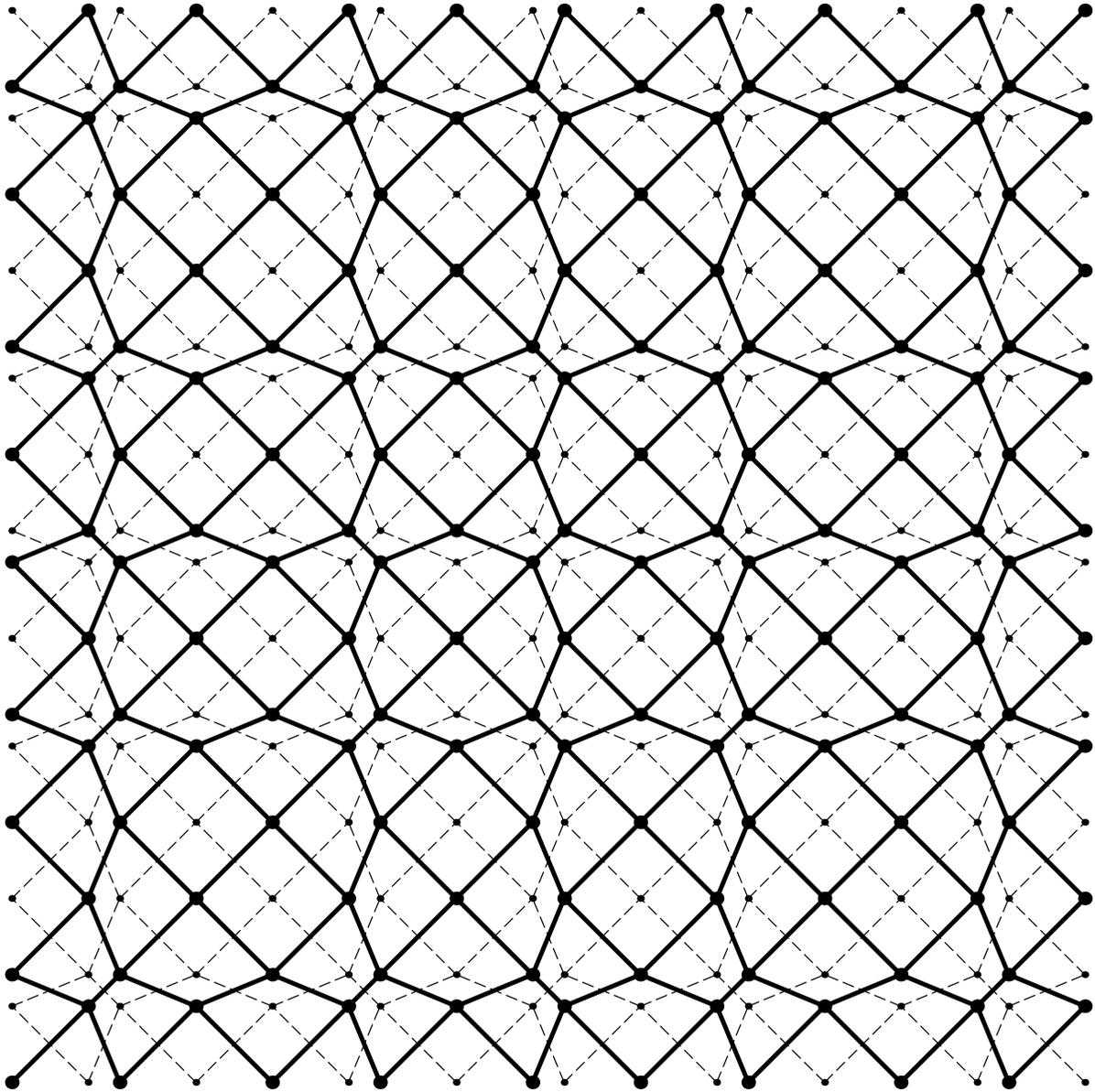}}
\caption{The Labyrinth and its dual}
\end{figure}\clearpage

\begin{figure}
\centerline{\epsfxsize=\textwidth\epsfbox[72 160 540 630]{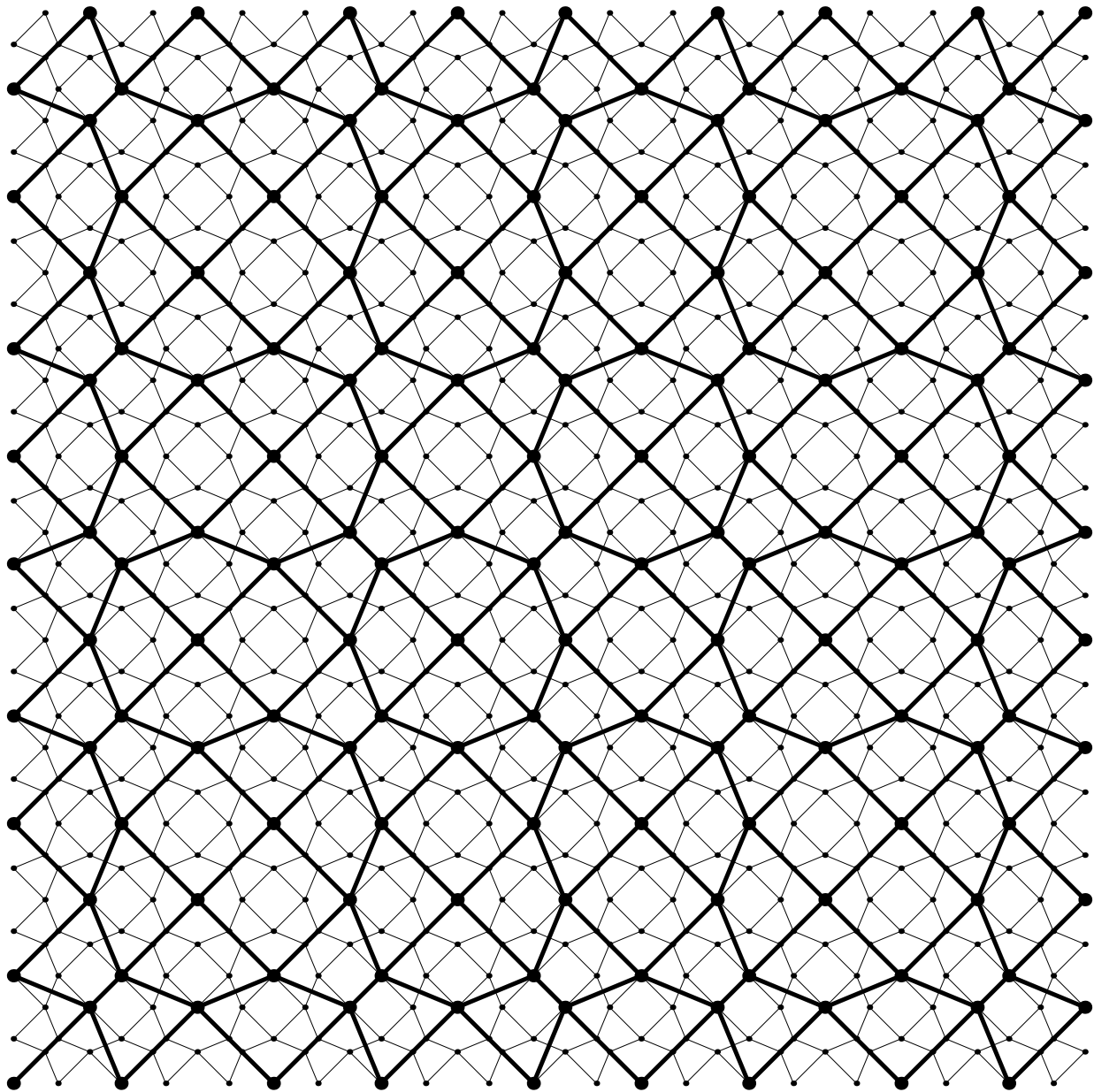}}
\caption{The tiling and its inflation}
\end{figure}\clearpage

\begin{figure}
\centerline{\epsfxsize=\textwidth\epsfbox[80 260 505 560]{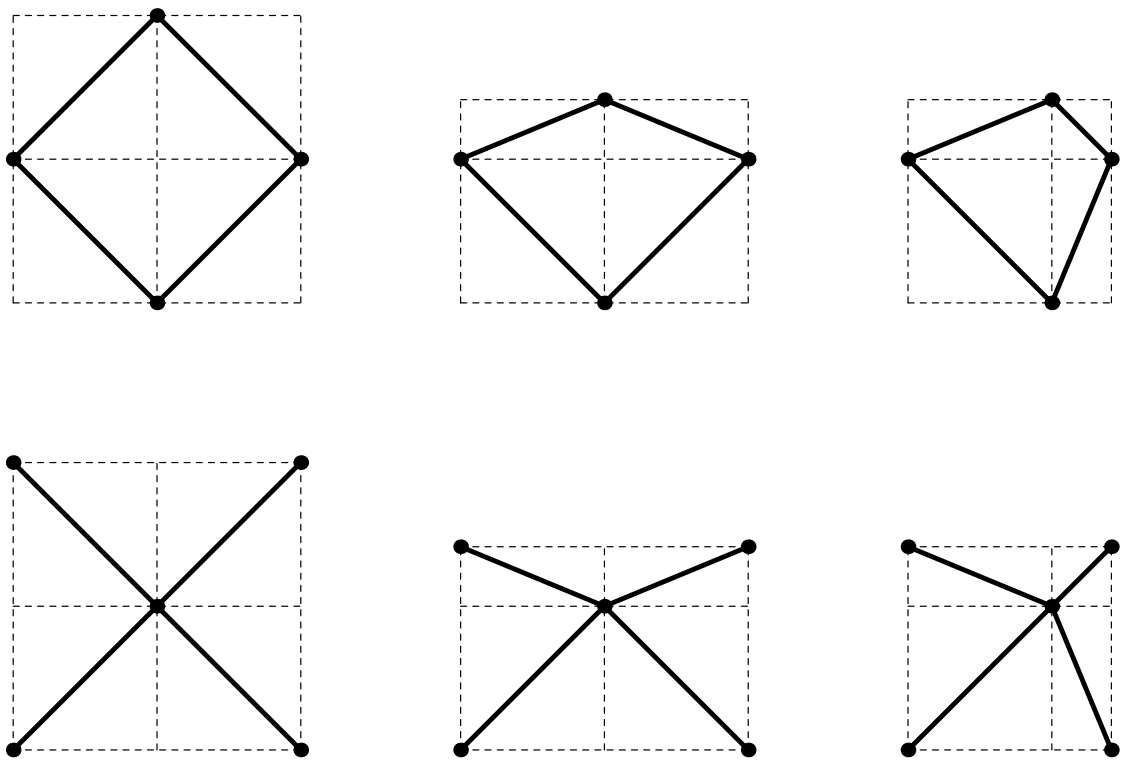}}
\caption{The tiles and their dual vertex configurations}
\end{figure}\clearpage

\begin{figure}
\centerline{\epsfxsize=\textwidth\epsfbox[72 120 540 700]{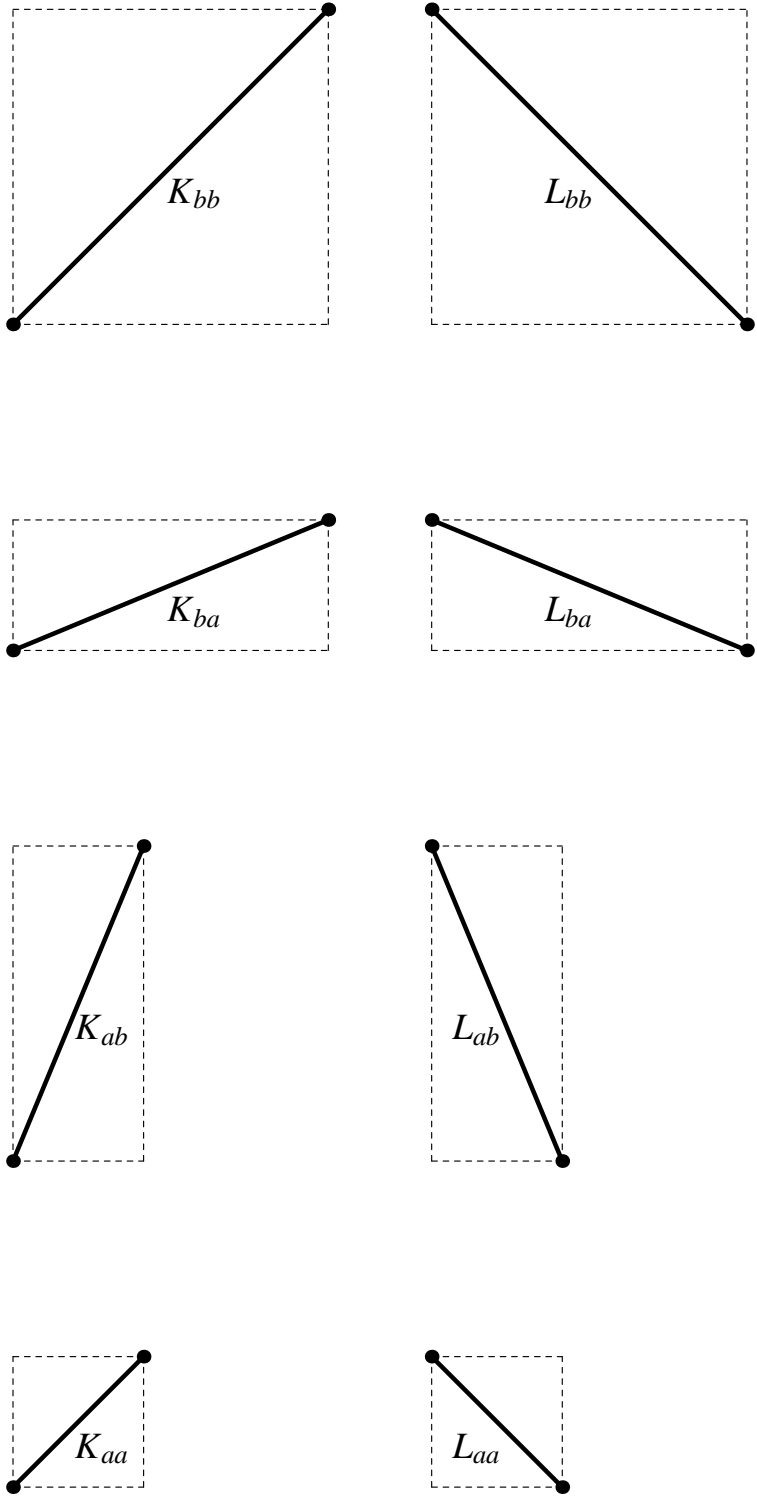}}
\caption{Assignment of couplings $K_{xy}$ and $L_{xy}$
        ($xy\in\{aa,ab,ba,bb\}$)}
\end{figure}\clearpage

\begin{figure}
\centerline{\epsfxsize=\textwidth\epsfbox[100 160 540 600]{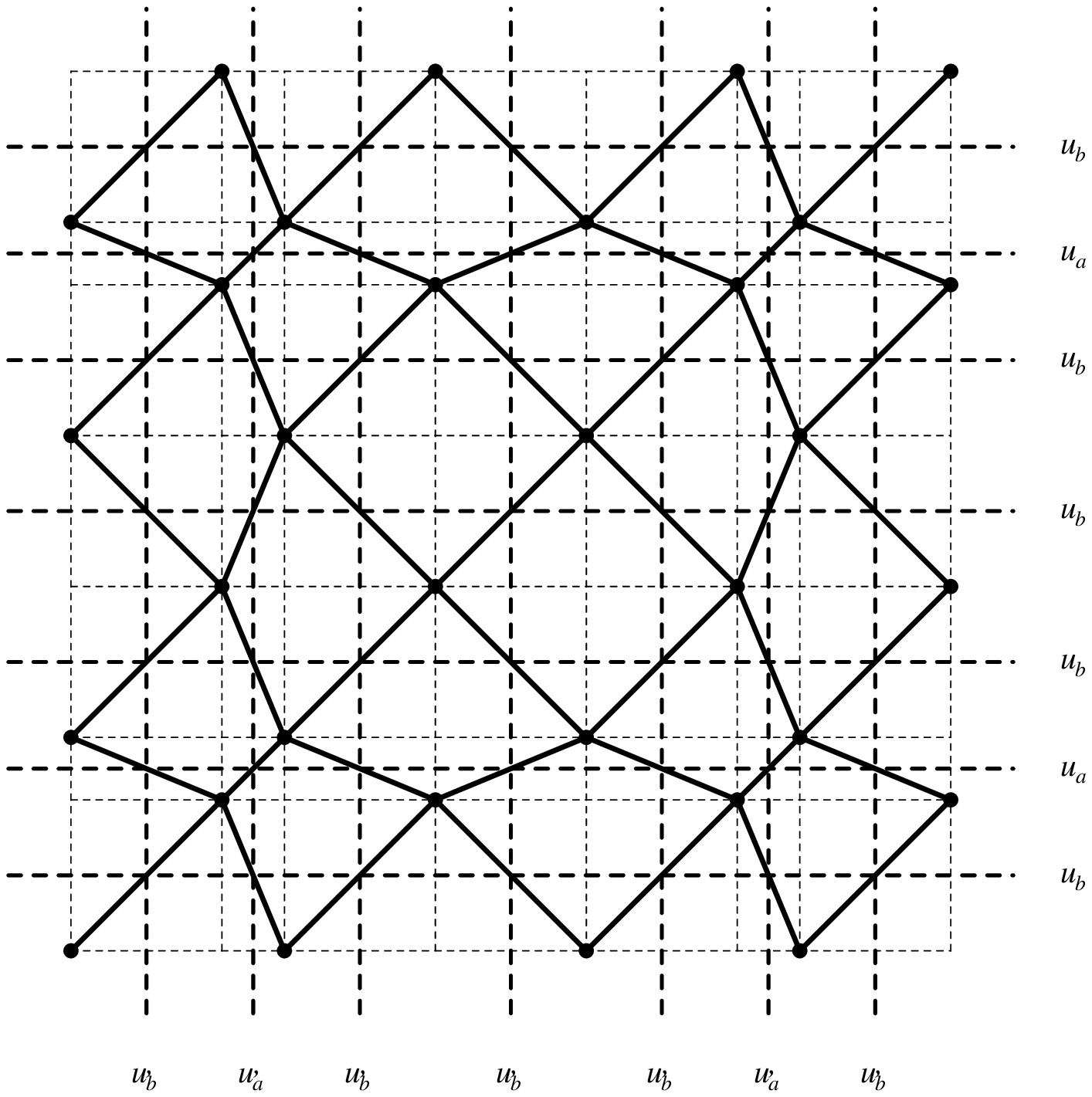}}
\caption{The tiling with rapidity lines and underlying grid}
\end{figure}


\begin{thebibliography}{99}

{\footnotesize

\bibitem{Cardy}
J.~L.~Cardy, 
\newblock {\em Conformal Invariance}, 
\newblock in: {\em Phase Transitions and Critical Phenomena}, vol.\ 11, 
\newblock eds.\ C.\ Domb und J.\ L.\ Lebowitz, 
\newblock Academic Press, London (1987).

\bibitem{Baxter}
R.~J.~Baxter,
\newblock {\em Exactly Solved Models in Statistical Mechanics},
\newblock Academic Press, London (1982).

\bibitem{VertexModels}
V.~Bazhanov,
\newblock {\em Trigonometric Solutions of Triangle Equations and
               Classical Lie Algebras},
\newblock \PLB{159}{85}{321};

M.~Jimbo,
\newblock {\em Quantum $R$ Matrix for the Generalized Toda System},
\newblock \CMP{102}{86}{537};

A.~Kuniba,
\newblock {\em Quantum $R$ matrix for G$_{2}$ and a solvable
               175-vertex model},
\newblock J.\ Phys.\ {\bf A23} (1990) 1349;

Z.-Q.~Ma,
\newblock {\em The spectrum-dependent solutions to the Yang-Baxter 
               equation for quantum E$_{6}$ and E$_{7}$},
\newblock J.\ Phys.\ {\bf A23} (1990) 5513;

J.~D.~Kim, I.~G.~Koh and Z.-Q.~Ma,
\newblock {\em Quantum $\check{R}$ matrix for $E_{7}$ and $F_{4}$ groups},
\newblock J.\ Math.\ Phys.\ {\bf 32} (1991) 845;

H.~J.~Chung and I.~G.~Koh,
\newblock {\em Solutions to the quantum Yang-Baxter equation for the
               exceptional Lie algebras with a spectral parameter},
\newblock J.\ Math.\ Phys.\ {\bf 32} (1991) 2406.

\bibitem{IRFModels}
G.~E.~Andrews, R.~J.~Baxter and P.~J.~Forrester,
\newblock {\em Eight-Vertex SOS Model and Generalized
           Rogers-Ramanujan-Type Identities}
\newblock \JSP{35}{84}{193};

\bibitem{Vladi}
G.~von Gehlen and V.~Rittenberg,
\newblock {\em $Z_n$-symmetric quantum chains with an infinite set
           of conserved charges and $Z_n$ zero modes},
\newblock \NPB{257}{85}{351}.

\bibitem{Baxter88}
R.~J.~Baxter,
\newblock {\em The superintegrable chiral Potts model},
\newblock \PLA{133}{88}{185}.

\bibitem{Kedem}
R.~Kedem and B.~M.~McCoy,
\newblock {\em Quasi-Particles in the Chiral Potts Model},
\newblock \IJMPB{8}{94}{3601}

\bibitem{BB93}
V.~V.~Bazhanov and R.~J.~Baxter,
\newblock {\em Star-Triangle Relation for a Three-Dimensional Model},
\newblock \JSP{71}{93}{839}.

V.~Pasquier,
\newblock {\em Two-dimensional critical systems labelled by
           Dynkin diagrams},
\newblock \NPB{285}{87}{162};

E.~Date, M.~Jimbo, A.~Kuniba, T.~Miwa and M.~Okado,
\newblock {\em Exactly solvable SOS models: Local height probabilities
           and theta function identities},
\newblock \NPB{290}{87}{231};

E.~Date, M.~Jimbo, A.~Kuniba, T.~Miwa and M.~Okado,
\newblock {\em Exactly solvable SOS models II: Proof of the
           star-triangle relation and combinatorical identities},
\newblock Adv.\ Stud.\ Pure Math.\ {\bf 16} (1988) 17;

M.~Jimbo, A.~Kuniba, T.~Miwa and M.~Okado,
\newblock {\em The $A_{n}^{(1)}$ Face Models},
\newblock \CMP{119}{88}{543};

M.~Jimbo, T.~Miwa, and M.~Okado,
\newblock {\em Solvable Lattice Models Related to the
               Vector Representation of Classical Simple
               Lie Algebras},
\newblock \CMP{116}{88}{507};

A.~Kuniba,
\newblock {\em Exact solution of solid-on-solid models for twisted
               affine Lie algebras $A_{2n}^{(2)}$ and $A_{2n-1}^{(2)}$},
\newblock \NPB{355}{91}{801};

A.~Kuniba and J.~Suzuki,
\newblock {\em Exactly solvable $G_{2}^{(1)}$ solid-on-solid models},
\newblock \PLA{160}{91}{216}.

\bibitem{DiluteModels}
S.~O.~Warnaar, B.~Nienhuis and K.~A.~Seaton,
\newblock {\em New Construction of Solvable Lattice Models Including
               an Ising Model in a Field},
\newblock \PRL{69}{92}{710};

S.~O.~Warnaar and B.~Nienhuis,
{\em Solvable lattice models labelled by Dynkin diagrams},
\newblock \JPhysA{26}{93}{2301};

S.~O.~Warnaar, B.~Nienhuis and K.~A.~Seaton,
\newblock {\em A Critical Ising Model in a Magnetic Field},
\newblock \IJMPB{7}{93}{3727};

S.~O.~Warnaar, P.~A.~Pearce, K.~A.~Seaton and B.~Nienhuis,
\newblock {\em Order Parameters of the Dilute A Models},
\newblock \JSP{74}{94}{469};

S.~O.~Warnaar, 
\newblock {\em Algebraic construction of higher rank dilute A models},
\newblock \NPB{435}{95}{463}.

\bibitem{BaaGriPis94}
M.~Baake, U.~Grimm and C.~Pisani,
\newblock {\em Partition Function Zeros for Aperiodic Systems},
\newblock \JSP{78}{95}{285}.

\bibitem{Benza89}
V.~G.~Benza,
\newblock {\em Quantum Ising Quasi-Crystal},
\newblock \EPL{8}{89}{321}.

\bibitem{LinTao92}
Z.~Lin and R.~Tao,
\newblock {\em Phase transition in aperiodic quantum Ising chains
           constructed with arbitrary substitution rules},
\newblock \PhysRev{B46}{92}{10808}.

\bibitem{Luck93}
J.~M.~Luck,
\newblock {\em Critical Behavior of the Aperiodic Quantum Ising
           Chain in a Transverse Magnetic Field},
\newblock \JSP{72}{93}{417}.

\bibitem{GriBaa94}
U.~Grimm and M.~Baake,
\newblock {\em Non-Periodic Ising Quantum Chains 
           and Conformal Invariance},
\newblock \JSP{74}{94}{1233}.

\bibitem{Tracy88}
C.~A.~Tracy,
\newblock {\em Universality classes of some aperiodic Ising models},
\newblock \JPhysA{21}{88}{L603}.

\bibitem{Igloi93}
F.~Igl\'{o}i,
\newblock {\em Critical behaviour in aperiodic systems},
\newblock \JPhysA{26}{93}{L703}.

\bibitem{Baxter78}
R.~J.~Baxter,
\newblock {\em Solvable eight-vertex model on an
           arbitrary planar lattice},
\newblock \PTRSL{289}{78}{315}.

\bibitem{Korepin}
V.~E.~Korepin,
\newblock {\em Eight-vertex model of the quasicrystal},
\newblock \PLA{118}{86}{285};

V.~E.~Korepin,
\newblock {\em Exactly solvable spin models in quasicrystals},
\newblock Sov.\ Phys.\ JETP {\bf 65} (1986) 614;

V.~E.~Korepin,
\newblock {\em Completely integrable models in quasicrystals},
\newblock \CMP{110}{87}{157}. 

\bibitem{Choy}
T.~C.~Choy, {\em Ising models on two-dimensional quasi-crystals: some
exact results}, Int.\ J.\ Mod.\ Phys.\ {\bf B2} (1988) 49.

\bibitem{deBruijn}
N.~G.~de Bruijn, {\em Algebraic theory of Penrose's non-periodic
tilings of the plane}, part I, Math.\ Proc.\ {\bf A84} (1981) 39-52,
and part II, Math.\ Proc.\ {\bf A84} (1981) 53-66.

\bibitem{GordLuckOrl86}
C.~Godr\`{e}che, J.-M.~Luck and H.~Orland, 
\newblock {\em Magnetic Phase Structure on the Penrose Lattice},
\newblock \JSP{45}{86}{777}.

\bibitem{AoyOda87}
H.~Aoyama and T.~Odagaki,
\newblock {\em Eight-Parameter Renormalization Group for
           Penrose Lattices},
\newblock \JSP{48}{87}{503}.

\bibitem{DoroSoka89}
A.~Doroba and K.~Sokalski,
\newblock {\em The Ising Model on the Two-Dimensional Quasiperiodic
           Lattice},
\newblock Physica stat.\ sol.\ {\bf B152} (1989) 275.

\bibitem{BhatHoJohn87}
S.~M.~Bhattacharjee, J.-S.~Ho and J.~A.~Y.~Johnson,
\newblock {\em Translational invariance in critical phenomena:
           Ising model on a quasi-lattice},
\newblock \JPhysA{20}{87}{4439}.

\bibitem{AmaAnaAth88}
G.~Amarenda, G.~Ananthakrishna and G.~Athithan,
\newblock {\em Critical Behavior of the Ising Model on a
           Two-Dimensional Penrose Lattice},
\newblock \EPL{5}{88}{181}.

\bibitem{OkaNii88a}
Y.~Okabe and K.~Niizeki,
\newblock {\em Monte Carlo Simulation of the Ising Model on the
           Penrose Lattice},
\newblock J.\ Phys.\ Soc.\ Japan {\bf 57} (1988) 16.

\bibitem{OkaNii88b}
Y.~Okabe and K.~Niizeki,
\newblock {\em Duality in the Ising model on quasicrystals},
\newblock J.\ Phys.\ Soc.\ Japan {\bf 57} (1988) 1536;

Y.~Okabe and K.~Niizeki,
\newblock {\em Phase transition of the Ising model on
           the two-dimensional quasicrystals},
\newblock J.\ Phys.\ Colloq.\ (France) {\bf 49} (1988) 1387.

\bibitem{WilVau88}
W.~G.~Wilson and C.~A.~Vause,
\newblock {\em Evidence for universality of the Potts model
           on the two-dimensional Penrose lattice},
\newblock \PLA{126}{88}{471}.

\bibitem{AbeDot89}
R.~Abe and T.~Dotera,
\newblock {\em High-Temperature Expansion for the Ising Model
           on the Penrose Lattice},
\newblock J.\ Phys.\ Soc.\ Japan {\bf 58} (1989) 3219;

T.~Dotera and R.~Abe,
\newblock {\em High-Temperature Expansion for the Ising Model
           on the Dual Penrose Lattice},
\newblock J.\ Phys.\ Soc.\ Japan {\bf 59} (1990) 2064.

\bibitem{Sor91}
E.~S.~S{\o}rensen, M.~V.~Jari\'c and M.~Ronchetti,
\newblock {\em Ising model on Penrose lattices: boundary conditions},
\newblock \PhysRev{B44}{91}{9271}.

\bibitem{SireMossSad89}
C.~Sire, R.~Mosseri and J.-F.~Sadoc,
\newblock {\em Geometric study of a 2D tiling related to
           the octagonal quasiperiodic tiling},
\newblock \JPhysFrance{50}{89}{3463}.

\bibitem{Peierls}
R.~Peierls,
\newblock {\em On Ising's model of ferromagnetism},
\newblock Proc.\ Cambridge Phil.\ Soc.\ {\bf 32} (1936) 477.

\bibitem{Griff}
R.~B.~Griffiths,
\newblock {\em Rigorous Results and Theorems},
\newblock in: {\em Phase Transitions and Critical Phenomena}, vol.~1,
\newblock eds.\ C.~Domb and M.~S.~Green,
\newblock Academic Press, London (1972).

\bibitem{Ellis}
R.~S.~Ellis,
\newblock {\em Entropy, Large Deviations, and Statistical Mechanics},
\newblock Springer, New York (1985).

\bibitem{3DIsing}
D.~W.~Wood,
\newblock {\em A self dual relation for a three dimensional assembly},
\newblock \JPhysC{5}{72}{L181};

P.~A.~Pearce and R.~J.~Baxter,
\newblock {\em Duality of the three-dimensional Ising model with
           quartet interactions},
\newblock \PhysRev{B24}{81}{5295};

R.~Liebmann,
\newblock {\em Monte Carlo Study of the Quartet Ising Model
           Consistent with Selfduality},
\newblock \ZPhys{B45}{82}{243}.

\bibitem{Baxter86}
R.~J.~Baxter
\newblock {\em Free-fermion, checkerboard and $Z$-invariant lattice models
           in statistical mechanics},
\newblock \PRSL{A404}{86}{1}.

\bibitem{AntKor88}
N.~V.~Antonov and V.~E.~Korepin,
\newblock {\em Critical properties of completely integrable
           spin models in quasicrystals},
\newblock Theor.\ Math.\ Phys.\ {\bf 77} (1988) 1282.

\bibitem{Doro89}
A.~Doroba,
\newblock {\em Equivalence of the Ising model on the 2-D Penrose 
           and 2-D regular lattices},
\newblock Acta Phys.\ Pol.\ {\bf A76} (1989) 949.

\bibitem{Sire89}
C.~Sire,
\newblock {\em Electronic Spectrum of a 2D Quasi-Crystal
            Related to the Octagonal Quasi-Periodic Tiling},
\newblock \EPL{10}{89}{483}.

\bibitem{BaaGriJos93}
M.~Baake, U.~Grimm and D.~Joseph,
\newblock {\em Trace maps, invariants, and some of their
           applications},
\newblock \IJMPB{7}{93}{1527}.

\bibitem{Katz}
A.~Katz,
\newblock {\em On the distinction between quasicrystals and
           modulated crystals}, 
\newblock in: {\em Quasicrystals}, 
\newblock eds.\ M.\ V.\ Jari\'c and S.\ Lundqvist, 
\newblock World Scientific, Singapore (1990), p.\ 200.

\bibitem{KraWa}
H.~A.~Kramers and G.~H.~Wannier,
\newblock {\em Statistics of the Two-Dimensional Ferromagnet. Part I},
\newblock Phys.\ Rev.\ {\bf 60}\ (1941) 252.

\bibitem{DuneDunlOgu93}
M.~Duneau, F.~Dunlop and C.~Oguey,
\newblock {\em Ground states of frustrated Ising quasicrystals},
\newblock \JPhysA{26}{93}{2791}.

\bibitem{Luck87}
J.~M.~Luck,
\newblock {\em Frustration effects in qusicrystals: an exactly
           soluble example in one dimension},
\newblock \JPhysA{20}{87}{1259}.

\bibitem{Sire93}
C.~Sire,
\newblock {\em Ising chain in a quasiperiodic magnetic field},
\newblock \IJMPB{7}{93}{1551}.

\bibitem{OitAydJohn90}
J.~Oitmaa, M.~Aydin, and M.~J.~Johnson,
\newblock {\em Antiferromagnetic Ising model on the Penrose lattice},
\newblock \JPhysA{23}{90}{4537}.

\bibitem{GradRyz65}
I.~S.~Gradshteyn and I.~M.~Ryzhik,
\newblock {\em Tables of Integrals, Series and Products}, 5th ed.,
\newblock Academic Press, London (1993).

}

\end{thebibliography}
\end{document}